\documentclass[
    ,final            
  ]
  {aipproc}

\layoutstyle{8x11double}

\begin{document}

\title{A Comparison of Gamma-ray Burst Subgroups Measured by RHESSI and BATSE}

\classification{01.30.Cc, 95.55.Ka, 95.85.Pw, 98.70.Rz}
\keywords      {gamma-ray astrophysics, gamma-ray bursts}

\author{Jakub \v{R}\'{i}pa}{address={Astronomical Institute of the Charles University, V Hole\v{s}ovi\v{c}k\'{a}ch 2, Prague, Czech Republic}}

\author{David Huja}{address={Astronomical Institute of the Charles University, V Hole\v{s}ovi\v{c}k\'{a}ch 2, Prague, Czech Republic}}

\author{Ren\'{e} Hudec}{address={Astronomical Institute, Academy of Sciences of the Czech Republic, Ond\v{r}ejov, Czech Republic}}

\author{Wojtek Hajdas}{address={Paul Scherrer Institut, Villigen, Switzerland}}

\author{Claudia Wigger}{address={Paul Scherrer Institut, Villigen, Switzerland},altaddress={Kantonsschule Wohlen, Switzerland}}

\begin{abstract}

A sample of almost 400 Gamma-ray bursts (GRBs) detected by the RHESSI satellite is studied statistically. We focus on GRB duration and hardness ratio and use the statistical $\chi^2$ test and the F-test to compare the number of GRB subgroups in the RHESSI database with that of the BATSE database. Although some previous articles based on the BATSE catalog claim the existence of an intermediate GRB subgroup, besides long and short, we have not found a statistically significant intermediate subgroup in the RHESSI data.

\end{abstract}

\maketitle

\section{RHESSI Instrument}

The Ramaty High Energy Solar Spectroscopic Imager (RHESSI) is a NASA Small Explorer satellite designed to study hard X-rays and gamma-rays from solar flares. It consists mainly of an imaging tube and a spectrometer. The spectrometer consists of nine germanium detectors (7.1 cm diameter and 8.5 cm height). They are lightly shielded only, thus making RHESSI also very useful to detect non solar photons from any direction. The energy range for GRB detection extends from about 50 keV up to 17 MeV depending on the direction. The energy and time resolutions are: $\Delta E$~=~3~keV (at 1000 keV), $\Delta t$~=~1~$\mu$~s. An effective area for near axis direction of incoming photons reaches up to 200 cm$^2$ at 200 keV. With a field of view of about half of the sky, RHESSI observes about one or two gamma-ray bursts per week.

\section{Data Samples}

Two samples are used. The first is the set of 358 GRBs observed by the RHESSI satellite and covers the period from February 2002 to April 2008 (see: \url{http://grb.web.psi.ch}). We have used the SSW program under IDL and authors' routines to derive count light-curves and count fluences. The second data-sample is the set of 1234 GRBs observed by the BATSE instrument. It is the data-set of 4B BATSE catalogue as is presented at \url{http://f64.nsstc.nasa.gov/batse/grb/catalog/4b/} This catalogue covers bursts detected up to 29 August 1996. Here we present a comparison of the statistical analysis of these two samples in order to show discrepancy of bursts with duration around 3 s.

\section{Duration Distribution}

\begin{figure}
  \includegraphics[width=0.75\textwidth]{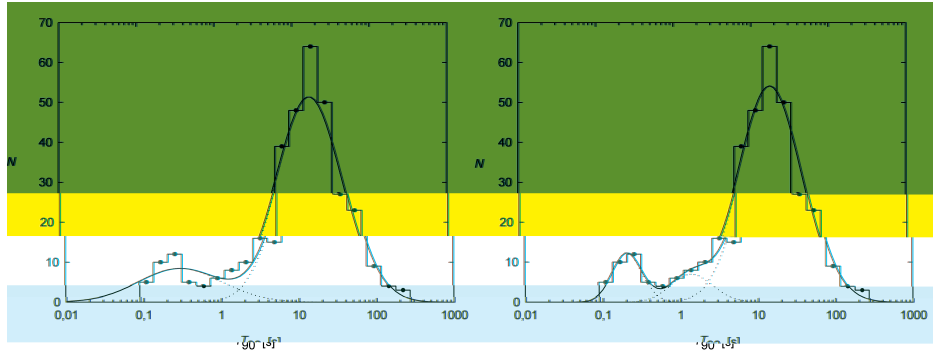}
  \caption{Distribution of the GRB RHESSI durations and 2 log-normal and 3 log-normal fits.}
  \label{RHESSI_durations}
\end{figure}

\begin{figure}
  \includegraphics[width=0.75\textwidth]{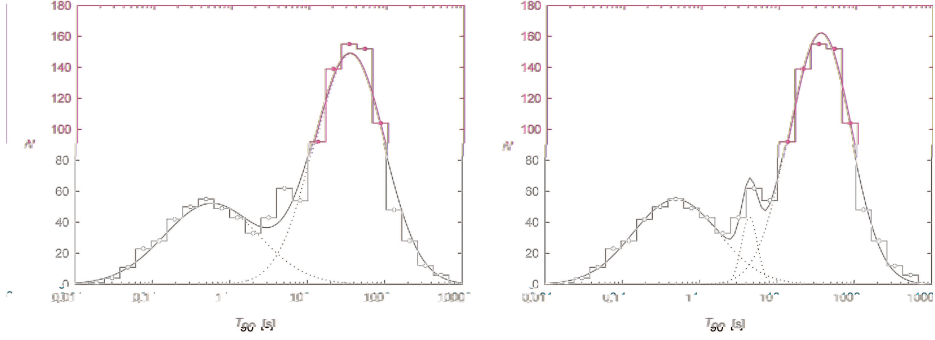}
  \caption{Distribution of the GRB BATSE durations and 2 log-normal and 3 log-normal fits.}
  \label{BATSE_durations}
\end{figure}

Here we present a distribution of durations $T_{90}$. Originally it was found (results from BATSE, Konus-Wind instruments \cite{1,2}) that there exist two subclasses. The short one with $T_{90}$ < 2 s and the long one with $T_{90}$ > 2 s. However, some articles point to existence of three subclasses of GRBs in the BATSE database with respect to their durations \cite{3,4}. Some articles also say that the third subclass (with intermediate duration), observed by BATSE, is a bias caused by an instrumental effect \cite{5}. Therefore, we decided to investigate this in the RHESSI database and compare the results with analysis of the BATSE data. We followed the method done in \cite{3} and fitted one, two and three log-normal functions (Figure~\ref{RHESSI_durations} for the RHESSI data and Figure~\ref{BATSE_durations} for the BATSE data) and we used $\chi^2$ test to evaluate these fits. The parameters of the fits, the $\chi^2$ values and goodness of the fits are noted in Table~\ref{table}. The question is whether the improvement of $\chi^2$ is statistically significant. To answer this question, we used F-test, as is described in \cite{6}. We obtained critical value $F_{0}$ which implies the probability $P$($F$>$F_{0}$). It is the probability that the improvement is just a statistical fluctuation.

\begin{center}
\begin{table} \caption{Parameters of 2 and 3 log-normal fits of $T_{90}$ of the RHESSI
                       and BATSE data-sets. $\mu$ are the means, $\sigma$ are standard
                       deviations and $w$ are the weights of the distribution.}
\centering
\begin{tabular}{lrrrr}

\hline
\\[-2ex]
parameter                    & RHESSI 2 fit     & RHESSI 3 fit      & BATSE 2 fit    & BATSE 3 fit  \\[0.3ex]
\hline\hline
\\[-2ex]
$\mu_{\textrm{short}}$       &          -0.52   &            -0.68 &  -0.24          & -0.31        \\[0.3ex]
$\sigma_{\textrm{short}}$    &          0.54    &             0.19 &   0.61          &  0.55        \\[0.3ex]
$w_{\textrm{short}}[\%]$     &          17.3    &            9.0   &   32.0          &  30.5        \\[0.3ex]
\hline
$\mu_{\textrm{long}}$        &          1.23    &             1.25 &   1.51          &  1.55        \\[0.3ex]
$\sigma_{\textrm{long}}$     &          0.42    &             0.41 &   0.46          &  0.40        \\[0.3ex]
$w_{\textrm{long}}[\%]$      &          82.7    &            83.8  &   68.0          &  65.1        \\[0.3ex]
\hline
$\mu_{\textrm{middle}}$      &                  &             0.15 &                 &  0.65        \\[0.3ex]
$\sigma_{\textrm{middle}}$   &                  &             0.27 &                 &  0.10        \\[0.3ex]
$w_{\textrm{middle}}[\%]$    &                  &            7.2   &                 &  4.4         \\[0.3ex]
\hline
$dof$                        &          14      &            11    &     18          &  15          \\[0.3ex]
$\chi^2$                     &          17.52   &            9.86  &   17.9          &  2.44        \\[0.3ex]
goodness[\%]                 &          22.9    &            54.3  &   46.2          & 99.99        \\[0.3ex]
$ F_{0} $                    &          2.85    &                  &   31.7          &              \\[0.3ex]
$P(F>F_{0})[\%] $            &           8.6    &                  & 10$^{-4}$       &              \\[0.3ex]
\\[-2ex]
\hline
\label{table}
\end{tabular}
\end{table}
\end{center}

\begin{figure}
  \includegraphics[height=0.21\textheight]{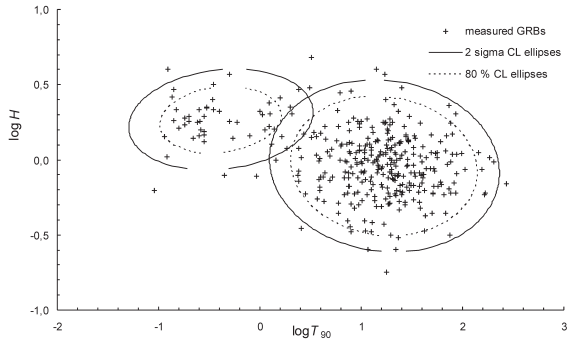}
  \caption{Hardness ratio vs. durations of the RHESSI GRBs with the best fit of two bivariate log-normal functions.}
  \label{RHESSI_hardness}
\end{figure}

\section{Hardness Ratio vs. Duration}

Mukherjee \cite{7} pointed to the existence of three GRB subclasses in multiparameter space. Articles \cite{8, 9} showed that in 2D plate of hardness ratio vs. $T_{90}$ of the BATSE data-set, three subclasses of GRBs can be found. The hardness ratio is defined as a ratio of two fluences F in two different energy bands integrated over the time interval $T_{90}$. They have mentioned that the third subclass has typical durations of about 4 s and the lowest hardness ratio which is anti-correlated with the duration. 

Here we present such distribution for the RHESSI data-set. Specifically we used hardness ratio as the ratio of fluences in the bands: (120 - 1500) keV / (25 - 120) keV. In Figure~\ref{RHESSI_hardness}, we show the RHESSI hardness ratio H vs. $T_{90}$ with the best fit of two bivariate log-normal functions. For the estimation of the best fit, we have used the maximum likelihood method (see references above and references therein). As one can see, the observed distribution can be sufficiently described by only two groups, and points outsides the CL ellipses do not make any lone compact cluster.

\section{Conclusion}

Despite a higher value of the fit's goodness with three log-normal functions on the $T_{90}$ histogram, in comparison with the fit of two log-normal curves, in the RHESSI data-set, F-test gives the probability $P$($F$>$F_{0}$) = 8.6 \% which is still too high to reject the hypothesis that the improvement in $\chi^2$ is only a statistical fluctuation.

Also the hardness ratio vs. duration plot  does not demonstrate any remarkable third subgroup (with intermediate duration and soft hardness as are present in the BATSE data-sample) in our RHESSI data. Therefore, we can not confirm the intermediate class of GRBs in the RHESSI sample.

Another result is the number of the GRBs with $T_{90}$ < 2 s which is for RHESSI $\simeq 14$ \% and for BATSE 4B $\simeq 26$ \%. Therefore, the ratio short/long GRBs obviously depends on the instrument, but this uncertainty does not change the conclusion that the short and long GRBs can be different phenomena \cite{10,11}.

\begin{theacknowledgments}

This study was supported by the GAUK grant No. 46307, by the Grant Agency of the Czech Republic, grant No. 205/08/H005, and by the Research Program MSM0021620860 of the Ministry of Education of the Czech Republic. The authors appreciate valuable discussions and help of A. M\'esz\'aros.

\end{theacknowledgments}


\begin{thebibliography}{9}

\bibitem{1}
  C.~Kouveliotou, et al., \emph{ApJ} \textbf{413}, 101 (1993).

\bibitem{2}
  R.~L.~Aptekar, et al., \emph{AIP Conf. Proc. of 4$^{th}$ Huntsville Symposium} p. 10 (1998).

\bibitem{3}
  I.~Horv\'{a}th, \emph{ApJ} \textbf{508}, 757 (1998).

\bibitem{4}
  I.~Horv\'{a}th, \emph{A\&A} \textbf{392}, 791 (2002).

\bibitem{5}
  J.~Hakkila, et al., \emph{ApJ} \textbf{538}, 165 (2000).

\bibitem{6}
  D.~L.~Band, et al., \emph{ApJ} \textbf{485}, 747 (1997), appendix A.

\bibitem{7}
  S.~Mukherjee, et al., \emph{ApJ} \textbf{508}, 314 (1998).

\bibitem{8}
  I.~Horv\'{a}th, et al., \emph{Baltic Astronomy} \textbf{13}, 217 (2004).

\bibitem{9}
  I.~Horv\'{a}th, et al., \emph{A\&A} \textbf{447}, 23 (2006).

\bibitem{10}
  L.G.~Bal\'{a}zs, et al., \emph{A\&A} \textbf{401}, 129 (2003).

\bibitem{11}
  L.G.~Bal\'{a}zs, et al., \emph{AIP Conf. Proc.} \textbf{662}, 137 (2003).

\end{thebibliography}
\end{document}